\documentclass[preprint,3p,times]{elsarticle} 

\usepackage{natbib}       
\usepackage{geometry}     
\usepackage{fleqn}        
\usepackage{graphicx}     

\usepackage{amsmath}      
\usepackage{amssymb}      
\usepackage{booktabs}     
\usepackage{xcolor}       
\usepackage[colorlinks=true,linkcolor=blue,citecolor=blue,urlcolor=blue]{hyperref} 

\newcommand{\beq}{\begin{eqnarray}}
\newcommand{\eeq}{\end{eqnarray}}

\begin{document}

\title{Quantum Fisher Information Reveals UV-IR Mixing in the Strange Metal}

\author[1]{David Ba\l{}ut}%
\author[1]{Xuefei Guo}
\author[1]{Niels de Vries}
\author[1]{Dipanjan Chaudhuri}
\author[2]{Barry Bradlyn}
\author[1]{Peter Abbamonte}
\author[2]{Philip W. Phillips\corref{cor1}}%
\ead{dimer@illinois.edu}

\cortext[cor1]{Corresponding author}
\affiliation[1]{organization={Materials Research Laboratory, University of Illinois at Urbana-Champaign},
addressline={},
city={Urbana},
state={IL},
postcode={61801},
country={USA}}
\affiliation[2]{organization={Department of Physics and Institute of Condensed Matter Theory, University of Illinois at Urbana-Champaign},
addressline={},
city={Urbana},
state={IL},
postcode={61801},
country={USA}}

\begin{abstract}

The density-density response in optimally doped Bi$_2$Sr$_2$CaCu$_2$O$_{8+x}$ has recently been shown to exhibit conformal symmetry.  Using, the experimentally inferred conformal dynamic susceptibility, we compute the resultant quantum Fisher information (QFI), a witness to multi-partite entanglement. For a Fermi liquid, we find that the QFI grows quadratically as the temperature increases,  consistent then with the phase space available for scattering in the standard theory of metals. By contrast, the QFI in a strange metal increases as a power law at as the temperature decreases, but ultimately extrapolates to a constant at $T=0$.  The constant is of the form, $\omega_g^{2\Delta}$, where $\Delta$ is the conformal dimension and $\omega_g$ is the UV cutoff which is on the order of the pseudogap. As this constant {depends on both UV and IR properties}, it illustrates that multipartite entanglement in a strange metal exhibits UV-IR mixing, a benchmark feature of doped Mott insulators as exemplified by dynamical spectral weight transfer.  We conclude with a discussion of the implication of our results for low-energy reductions of the Hubbard model.

\end{abstract}

\maketitle

\section{Introduction}

In championing that strong quantum entanglement and criticality underlie the physics of the strange metal, Jan Zaanen\cite{zaanenscipost} adopted the moniker ``many body entangled compressible quantum matter.''  While by entanglement, Zaanen meant the standard deviation from a tensor product state entailed by  linear combination of two seemingly contradictory states, also known as `cat' states, the type of quantum criticality he had in mind was more nuanced. In typical quantum critical scenarios with a single diverging length scale, $\xi\propto(g-g_c)^{-\nu}$, one also specifies that the divergence of time correlations is dictated by the spatial correlations, namely, $\xi_\tau\propto \xi^z$, where $z$ is the dynamical exponent.  What Zaanen had in mind is a quantum critical or more accurately a conformally invariant scenario in which the canonical spatial coordinates play no role.  Consider a charged Reissner-Nordstrom black hole in a bulk AdS$_4$ background. The near horizon geometry dictates that the IR physics has an emergent  AdS$_2\times R^2$ symmetry\cite{faulkner1,faulkner2}.   As AdS$_2$ possesses just a time and a radial coordinate, the renormalization direction, the matter that this describes is strongly correlated only temporally with no spatial extent, such that $z\rightarrow \infty$.  Such local quantum criticality would seem to be only a fiction if it were not for the experiments on the  density-density response function measured recently on BSCCO \cite{guo2024conformallyinvariantchargefluctuations}.  At low frequencies, the imaginary part of the density correlator exhibits only a weak momentum dependence, $B(q)$, but a frequency structure\cite{mitranoAnomalousDensityFluctuations2018,guo2024conformallyinvariantchargefluctuations}, 
\beq
    \chi''_{\rm BSCCO}(\omega,T) = B(q) T^{2 \Delta - 1}{\rm Im}\left[\frac{\Gamma\left(\Delta - i\frac{\hbar\omega}{2\pi k_BT}\right)}{\Gamma\left(1 - \Delta - i\frac{\hbar\omega}{2\pi k_BT}\right)}\right]
    \label{cftf},
\eeq
consistent with local quantum critical AdS$_2\times R^2$  conformal matter\cite{faulkner1}.  Here $\Delta$ is the conformal dimension which to match the experiments requires $\Delta=0.05$ and $B(q)$ is a weak momentum-dependent function. The separability of the space and frequency dependence mirrors the separability of space and time in AdS$_2\times R^2$.  Up to about $\omega_g=200$ meV, this expression fits the data extremely well.  Above $\omega\approx 200$ meV, $\chi{''}(\omega,T)$ is featureless and approximately temperature independent to 1 eV and then vanishes algebraically with frequency beyond this scale \cite{mitranoAnomalousDensityFluctuations2018,guo2024conformallyinvariantchargefluctuations}.  The algebraic fall-off at large frequency ensures congruence with the f-sum rule.   It turns out the minimal statement necessary to yield Eq. (\ref{cftf}) is simply power-law dependence in imaginary time for the two-point function, be it density or otherwise, coupled with a conformal mapping between the imaginary time axis with width $1/T$ to the infinite plane through  $w(\tau)=\exp (2\pi iT\tau)$. Conformal maps  consist of all angle-preserving distortions, stretching, compression, which preserve angles of lines and curves in a spacetime. Such symmetry is manifest in impurity problems, notably Kondo lattices \cite{Tsvelik1997}. The distinct peak structure away from $\omega=0$ ($\Delta=0$) in Eq. (\ref{cftf}) is absent if the conformal dimension were $\Delta=0.5$.   This limit is more akin to marginal Fermi liquid phenomenology\cite{mfl} in which momentum dependence is also absent\cite{mfl}. The breakdown of the MFL description in the experimentally-measured low-frequency limit of the density-density response is more than just an exponent problem as MFL posits a flat frequency dependence set by the temperature scale, quite different from what Eq. (\ref{cftf}) entails.

The consilience of the experiments with a locally critical conformal structure poses an overwhelming question:  What do doped 2D Mott insulators such as the cuprates have anything to do with conformally invariant local quantum criticality (LQC)? Aside from the dimensionality being wrong, doped Mott insulators possess a scale, namely the distance between the upper and lower Hubbard bands.   Doesn't the presence of this scale thwart any conformally invariant reduction?  Regarding this question, it might be instructive to probe the thermodynamics to see if there is any hint of LQC.  Within a restricted analysis, Ref.~\cite{legros} found a relationship between the coefficient of T-linear resistivity in the strange metal and the entropy.  With an analogy with the universal inequality\cite{etas} of the viscosity, $\eta$, to entropy density $s$, $\eta/s\ge \hbar/ 4\pi$, Zaanen\cite{zaanenscipost} posited a similar minimal viscosity principle for the cuprates.  An essential part of the argument is the fact that the entropy in a conformal theory scales as $S\propto T^{d/z}$.  In  LQC ($z=\infty$), no temperature dependence survives for the entropy in contrast to the experimental observations\cite{legros}.  A possibility\cite{zaanenscipost} is that the hyperscaling violation exponent effectively reduces the dimensionality to $d-\theta$ with the caveat that $-\theta\rightarrow\infty$ along with $z\rightarrow\infty$ but the ratio $-\theta/z$ remaining finite.  This procedure does yield a non-vanishing temperature dependence for the entropy even in the ultra-local limit in which $z\rightarrow\infty$. Hence, temperature dependence of the entropy in the ultra-local limit requires fine-tuning of the dynamical and hyperscaling exponents. 

What about the Mott scale?  As the presence of the Mott scale is undeniable in the cuprates,  a secondary question is does this play any role in the quantum entanglement?  If strong entanglement obtains, then it is difficult to disentagle purely lower from upper Hubbard band physics. Consequently, the entanglement should be linked to an energy scale. Identifying this energy scale is our focus here.  Entanglement entropy computations\cite{srednicki} provide information regarding the inherently quantum correlations between two subsystems obtained by  bipartition. 
If two regions are correlated, there is no reason to stop there.  For strongly correlated systems, a probe of multipartite correlations is a better probe of the entanglement.  It would seem that obtaining such knowledge\cite{Guo_2022,multi1} requires inherently non-local information about the system. In this context, the quantum Fisher information (QFI)\cite{Toth,hyllus,schiewitness,witness2}, $F_Q$, is a highly useful witness of multi-partite entanglement.   Developed in the context of metrology, the QFI is designed to distinguish how distinct an initial density matrix is from its value at a later time.  Consider a thermal mixed state with density matrix $\rho=\sum_\lambda p_\lambda|\rangle\langle\lambda|$ and $p_\lambda$ the Boltzmann weight for state $|\lambda\rangle$ with eigenenergy $E_\lambda$. The corresponding QFI\cite{fisher,bures},
\beq
F_Q=2\sum_{\lambda,\lambda'}\frac{(p_\lambda-p_{\lambda'})^2}{p_\lambda+p_{\lambda'}}\langle\lambda|\hat{O}|\lambda'\rangle|^2,
\label{fqizoller}
\eeq
just involves the matrix elements of the operator $\hat O$ which couple to an external perturbation (note that terms with $p_\lambda=p_{\lambda'}=0$ are excluded from the summation).  For an N-particle system, if $F_Q/N\equiv f_Q>m$, then the system is at least (m+1)-partite entangled.  A divergence of this quantity would imply then that all parts of the system are entangled.  Hence, the QFI is a measure of the total entanglement whether the degrees of freedom lie at low energy or not. In principle, the QFI should be sensitive to the high energy scales if such scales partake in the m+1-partite entanglement.  In addition, the QFI provides a way of defining the Bures\cite{bures} metric, $d_B^2(\hat{\rho},\hat{\rho}+d\hat{\rho})\propto F_Q dt^2$ for the evolution of the density matrix.  For pure states, the Bures metric reduces to the (Fubini-Study) quantum metric~\cite{provost1980riemannian}.

We show in this paper is that the QFI associated to density response, $\hat{O}=\rho_\mathbf{q}$, provides a unique measure of the entangled energy scales in the strange metal.  Using Eq. (\ref{cftf}), we compute $F_Q$ directly.  We find that the zero-temperature limit of $F_Q$ scales as $\omega_g^{2\Delta}$ and hence is dictated by both UV and IR physics.  Coupled with the fact that $F_Q\propto T^{-\Delta_Q/z}$ for critical systems~\cite{haukeMeasuringMultipartiteEntanglement2016}, our findings are consistent as $T\rightarrow 0$ only if $z=\infty$.  Hence, the QFI provides independent evidence for LQC. Further, the mixing of UV and IR physics in the QFI is not unexpected in a doped Mott insulator because removing a single electron rearranges the spectrum on all energy scales giving rise to static and dynamic spectral weight transfer\cite{sawatzky,phillipsrmp}. This has implications for any low-energy effective theory of doped Mott insulators. 

\section{Quantum Fisher Information}

 Recent inelastic neutron experiments have used the QFI to show multipartite spin entanglement in the  quasi-one-dimensional antiferromagnetic spin chain KCuF3\cite{schiewitness}.  At lower temperatures, the spin chain was shown to exhibit at least bipartite entanglement. Additionally, inelastic X-ray scattering was recently used to measure the QFI associated to density response in the insulator LiF~\cite{balutQuantumEntanglementQuantum2024}.  In the context of strange metals, Ref.~\cite{mazza,si2} computed the QFI from neutron spin susceptibility data on the heavy fermion compound Ce$_3$Pd$_20$Si$_6$ and saw a marked increase with no particular scale.  Motivated by such analyses, we turned our attention to optimally-doped BSCCO as there has been no investigation thus far of a witness of entanglement in the cuprates.  Operationally\cite{haukeMeasuringMultipartiteEntanglement2016}, Eq. (\ref{fqizoller}) can be rewritten,
\begin{align}
\label{eq:qfidef}
    F_Q(T)/N&\equiv f_Q(T)\nonumber \\
    &= \frac{4\hbar}{\pi}\int_0^\infty d\omega\;\mathrm{tanh}\left(\frac{\hbar\omega}{2k_BT}\right)\chi''(\mathbf{q},\omega,T),
\end{align}
directly in terms of the (intensive) dynamical susceptibility, $\chi^{''}(\mathbf{q},\omega,T)$, associated with perturbations by $\hat{O}=\rho_\mathbf{q}$, thereby making it amenable to direct experimental detection. This form for $F_Q$ leads  naturally to a scaling analysis near an underlying critical point.  Defining the distance to the critical point as ${\tilde g}$ and scaling all the lengths by temperature according to $L\propto 1/T^z$, results in the universal\cite{haukeMeasuringMultipartiteEntanglement2016} scaling, 
\beq
    f_{Q}(T,L^{-1},\tilde{g}) = T^{-\Delta_{Q}/z} \phi(L^{-1}T^{-1/z},\tilde{g}T^{-1/(z\nu)}),
    \label{eq: qfiScaling}
\eeq
for the QFI in the thermally dominated regime where if $L^{-1} < T^{1/z}$.  Here $\Delta_{Q}$ is defined as $\Delta_{Q} = d - 2\Delta_{\hat O}$ and $\Delta_{\hat O}$ is the scaling dimension of $\hat O$, which for the remainder of this work we will take to be the density operator.  We see then that the QFI density diverges if $\Delta_{Q}>0$ and asymptotes to a constant in the opposite regime.  This analysis presumes of course that the dynamical exponent is finite.  The situation of interest here is LQC in which $z=\infty$.

 \section{Lindhard Analysis: Free Electrons}

\begin{figure}
    \centering
    \includegraphics[width=\linewidth]{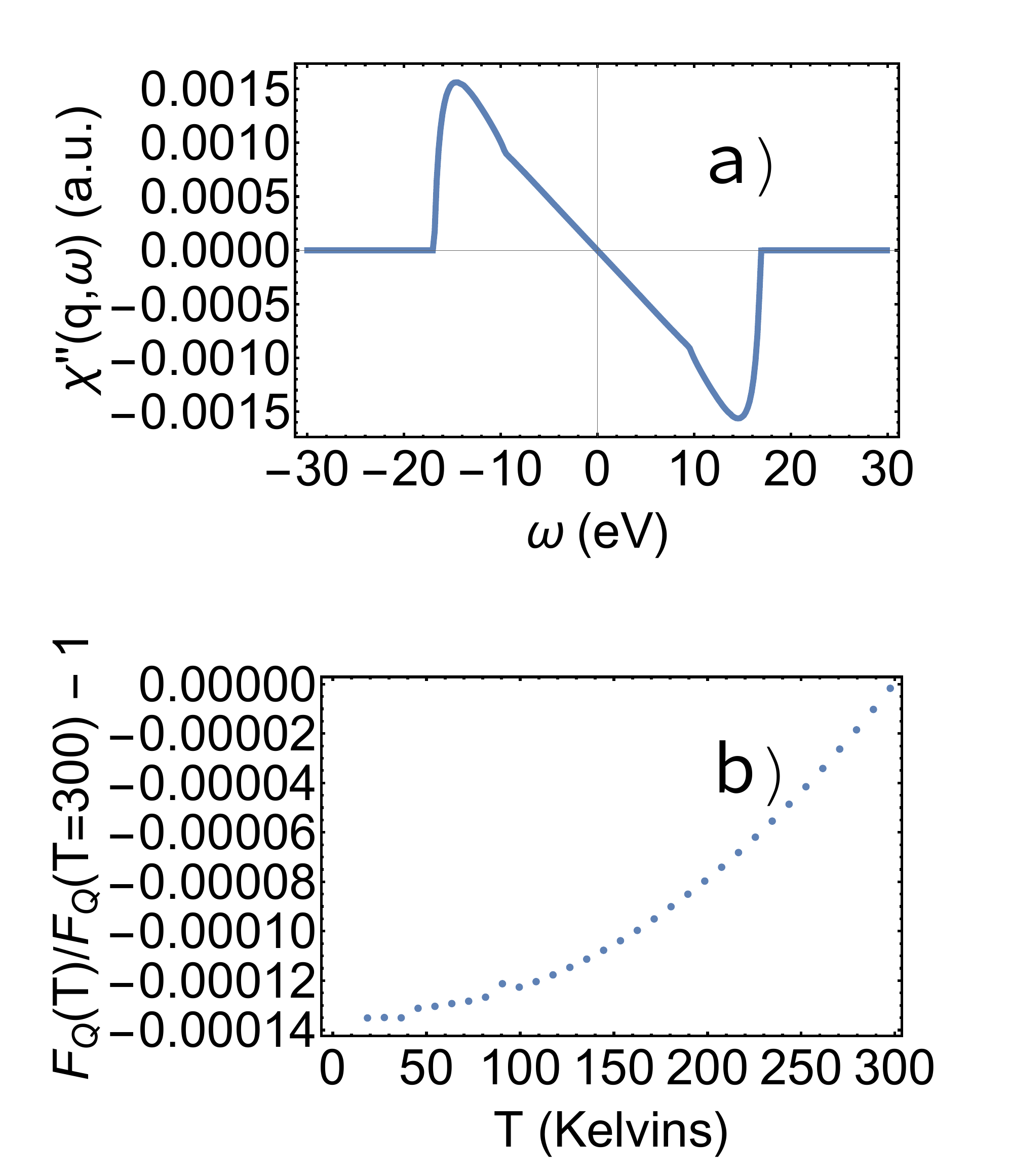}
    \caption{RPA response.  a) Dynamic RPA susceptibility for $q=0.55 k_F$ showing distinctly that it is an odd function of frequency.  b) Resultant QFI from RPA at $q=0.55k_F$.  The saturation at low temperatures is the signature of a weakly entangled state of matter.}
    \label{fig1}
\end{figure}

As a point of contrast, we first examine the QFI in an ordinary metal. We start with the Lindhard screening function (polarizability) $\Pi_0$ for free electrons. In terms of $q = k/k_{F}$ and $\nu = \frac{\omega m }{\hbar k_{F}^2}$, the real and imaginary parts of $\Pi_{0}(q,\nu)$ {at zero temperature} are given by\cite{FW}
\begingroup
\begin{align}
    \text{Re}&\Pi_{0}(q,\nu) = \frac{2m k_{F}}{4 \pi^2 \hbar^2 c^2} \Bigg(
    -1 + \frac{1}{2q}\Bigg[\left(1 - \left(\frac{\nu}{q} - \frac{q}{2}\right)^2\right) \nonumber \\
    &\quad\times\ln\left|\frac{1 + (\nu/q - q/2)}{1 - (\nu/q - q/2)}\right| 
    -\left(1 - \left(\frac{\nu}{q} + \frac{q}{2}\right)^2\right) \nonumber \\
    &\quad\times\ln\left|\frac{1 + (\nu/q + q/2)}{1 - (\nu/q + q/2)}\right|
    \Bigg]\Bigg)
\end{align}
\endgroup
and 
\begingroup
\begin{align}
    &\text{Im}\Pi_0(q, \nu) = \nonumber \\
    &\begin{cases}
        -\frac{mc^2 k_F}{\hbar^2 c^2} \frac{1}{4\pi q} \left(1 - \left(\frac{\nu}{q} - \frac{q}{2}\right)^2\right), \nonumber \\
        \quad \text{if } q > 2 \text{ and } \frac{q^2}{2} - q < \nu < \frac{q^2}{2} + q \\[1ex]
        -\frac{mc^2 k_F}{\hbar^2 c^2} \frac{1}{4\pi q} \left(1 - \left(\frac{\nu}{q} - \frac{q}{2}\right)^2\right), \nonumber \\
        \quad \text{if } q < 2 \text{ and } q - \frac{q^2}{2} < \nu < \frac{q^2}{2} + q \\[1ex]
        -\frac{mc^2 k_F}{\hbar^2 c^2} \frac{1}{4\pi q} 2\nu, \nonumber \\
        \quad \text{if } q < 2 \text{ and } 0 < \nu < q - \frac{q^2}{2} \\[1ex]
        0, \quad \text{otherwise}.
    \end{cases}
\end{align}
\endgroup
The generalization of $\Pi_0(q,\nu)$ to nonzero temperature can be similarly computed in closed form; we refer the reader to Chapter 9 of Ref.~\cite{phillipsbook} for the explicit expression. We use the polarizability to compute the density-density response of the interacting electron gas within the random phase approximation (RPA),
\beq
    \chi^{RPA}(q,\nu) = \frac{\Pi_{0}(q,\nu)}{1 - V(q) \Pi_{0}(q,\nu)},
\eeq
from the bubble summation with the 3-dimensional Coulomb interaction, $V(q)$.   Displayed in Fig.~\ref{fig1}a is precisely the RPA susceptibility for $q=0.55k_F$ and $T=0$ which exhibits the characteristic antisymmetry with respect to $\omega\rightarrow-\omega$. Using the temperature-dependent $\chi^{RPA}(q,\nu)$, we compute the QFI as a function of temperature  shown in Fig.~\ref{fig1}b.  Regardless of the momentum chosen, we find that the the QFI decreases and ultimately saturates at a small value at low temperature. This indicates that Fermi liquids are weakly entangled at $T=0$.   In fact, the normalized QFI we find here is only weakly dependent on temperature showing only a modest decrease of $.00012$ from the room-temperature value.  The constant is determined by the finite energy transfer at $T=0$.  This modest decrease is due to the fact that for the RPA response, the particle-hole continuum---rather than the plasma response---dominates the low-energy spectral weight.  Recall, the imaginary part of the susceptibility vanishes above a certain frequency while it is the real part that has a pole. From this analysis, we find that Fermi liquids display little to no multipartite entanglement at low temperature.  In the high-temperature limit, the QFI has to match on with the $f$-sum rule, $f_Q(T)\sim n_e (\hbar q)^2/mT$ signifying that density perturbations at wavevector q can excite almost every electron in the system with approximately equal probability. The fact that $f_Q$ decreases as $T\rightarrow 0$ and as $T\rightarrow\infty$ implies that a peak must exist at a nonzero temperature, which for a Fermi liquid must be on the order of the Fermi temperature.

We also note that the QFI as a measure of how sensitive the system is to density perturbations should be governed by the phase space for scattering.  For a Fermi liquid, the phase space should scale as $a+bT^2$.  This is precisely the temperature dependence of the QFI we see in Fig. (\ref{fig1}). The constant, $a$ is set by the finite frequency scattering at $T=0$.   Consequently, our findings here are consistent with the predicted behaviour for a Fermi liquid.
 For completeness, we also computed the QFI for a cylindrical Fermi surface\cite{Mihaila_2011}.  We noticed no substantial differences with the ratio $F_Q(T)/F_Q(T=300)-1=0.0005$, identical to that appearing in Fig. (\ref{fig1}).  

\section{Strange Metal Entanglement}

\begin{figure}
    \centering
    \includegraphics[width=\linewidth]{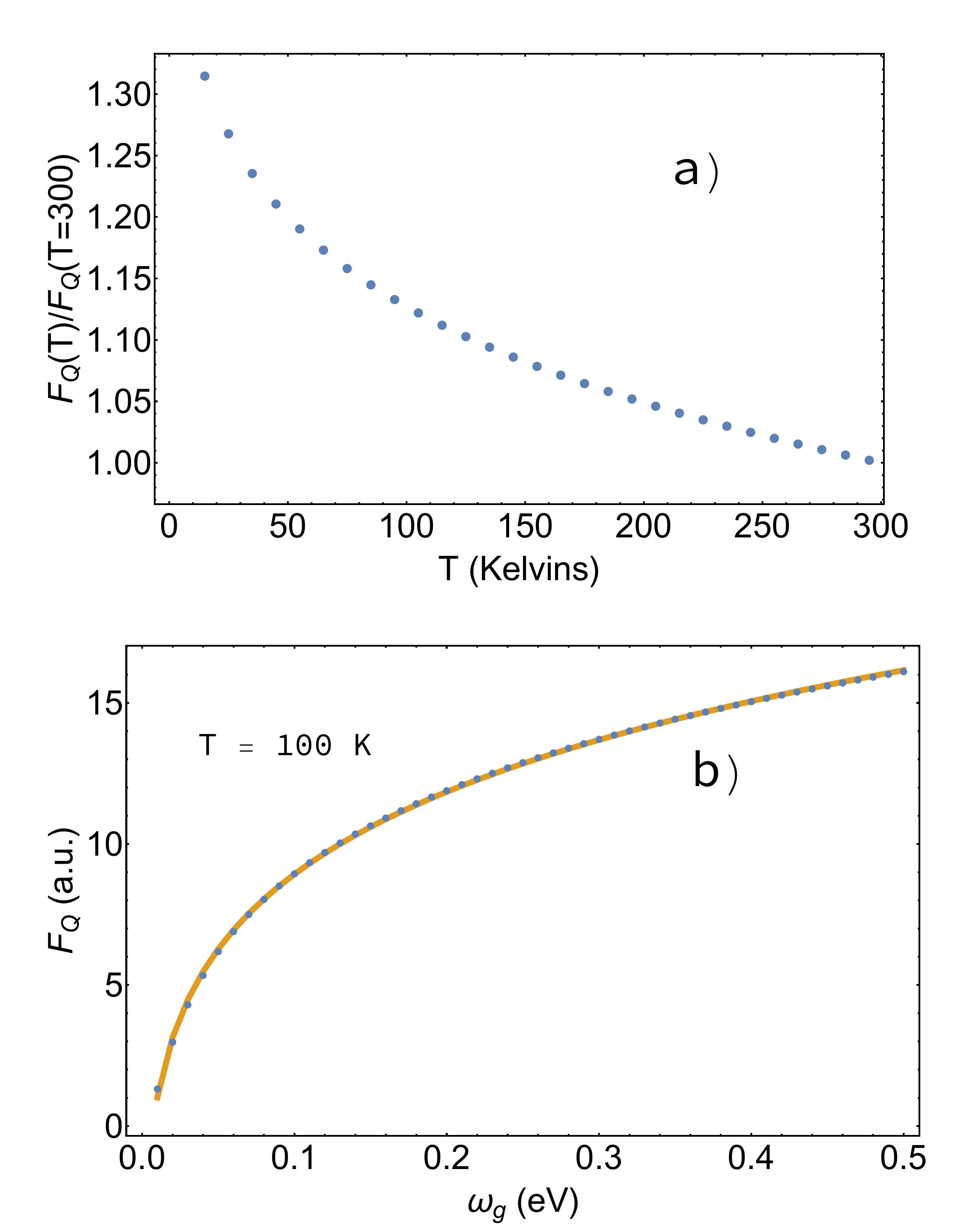}
    \caption{QFI for the strange metal.  a) QFI computed from Eq. (\ref{eq:qfidef}).  b) Fit of the $\omega_g$ dependence (measured in eV) of the QFI to a functional form $f_Q=c+a\omega_g^{2\Delta}$ for $\Delta=0.05$ and $T=100K$. The fit perameters were found to be $a\approx40,c\approx-20.2$ (note that the background $\Lambda$ was not included in the fit). The fit is consistent with the prediction of Eq.~(\ref{uvir}).}
    \label{fig2}
\end{figure}

Beyond RPA, we now turn to the conformally invariant dynamic susceptibility inferred from the experiments on BSCCO\cite{guo2024conformallyinvariantchargefluctuations}.  In the context of the experiments, we write the dynamic susceptibility as 
\begin{equation}
\chi_{\rm surface}''(T) = 
\begin{cases}\chi^{''}_{\rm BSCCO}(\omega, T) & \text{if } \omega < \omega_g \\
    \text{T-Independent spectrum} & \text{if } \omega > \omega_g.
\end{cases}
\label{chiexp}
\end{equation}
The constant overall scale of $B(q)$ in Eq.~\eqref{cftf} is found by imposing continuity with the continuum value at $\omega_g\approx 200 \ meV$.  To analyze the QFI, we substitute Eq. (\ref{chiexp}) into Eq. (\ref{eq:qfidef}) and perform the integration numerically.  Fig.~\ref{fig2}a shows a representative result.  The QFI is in general considerably larger than that obtained from the RPA response and follows a power-law increase with decreasing temperature, thereby underscoring the importance of multi-partite entanglement in a strange metal, in contrast to a Fermi liquid. Such an increase of the QFI has been noted previously on the heavy fermion Ce$_3$Pd$_{20}$Si$_{6}$ and CeCu$_{0.9}$Au$_{0.1}$based on an analysis\cite{mazza,si2} of the spin dynamic susceptibility. These results raise the possibility that the QFI in the strange metal diverges at low temperature.   Hence, analytics at low temperature are imperative.
It turns out we can extract the low-temperature behavior exactly.  To facilitate this analysis, we rewrite Eq. (\ref{cftf}) in a more tractable form. First, we define $x=\hbar\omega/2\pi k_BT$.  Noting that $\Gamma(1-\varsigma)\Gamma( \varsigma)=\pi/\sin\pi \varsigma$, we rewrite $1/\Gamma(1-\Delta-ix)=\Gamma(\Delta+ix)\sin(\pi(\Delta+ix)/\pi$.  Consequently, 
\beq
{\rm Im} \frac{\Gamma(\Delta-ix)}{\Gamma(1-\Delta-ix)}=\frac{|\Gamma(\Delta-ix)|^2}{\pi}\cos\Delta\sinh x.
\eeq
We are particularly interested in the low-temperature behaviour of the QFI.  To this end, we use Stirling's approximation,
\beq
\Gamma(\varsigma)\approx e^{-\varsigma} \varsigma^{\varsigma-1/2},
\eeq
for large $\zeta$ as we are interested in the low-temperature or large $x$ regime.
From this expression, we find that  
\beq
|\Gamma(\Delta+ix)|^2\approx e^{-2\Delta} e^{-\pi x} x^{2\Delta-1}
\eeq
acquires a particularly simple form when $x$ is large. The $e^{-\pi x}$ factor exactly cancels the leading behaviour of $\sinh x\approx e^{\pi x}$. Consequently, the QFI reduces to a simple algebraic integral,
\beq
F_Q&\approx& \Lambda+T^{2\Delta} e^{2\Delta}\frac{\cos\Delta}{2}\int_0^{\frac{\hbar\omega_g}{2\pi k_B T}} x^{2\Delta-1}dx\nonumber\\
&=&\Lambda+ \tilde C\left(\frac{\hbar\omega_g}{4\pi k_B}\right)^{2\Delta},
\label{uvir}
\eeq
in which the temperature dependence vanishes to leading order.   Here $\tilde C=B(q)e^{2\Delta}\cos\Delta/4\Delta$ and $\Lambda$ arises from the flat part of the dynamic susceptibility. Fig.~\ref{fig2}b confirms the low-temperature $\omega^{2\Delta}$ dependence. Whether the same behaviour holds for the QFI in heavy fermions\cite{mazza,si2} based on the spin susceptibility remains an open question. Consequently, we find that the zero-temperature QFI is a constant depending on the UV cutoff raised to the IR content, namely the conformal dimension, $\Delta$.   This is a surprising result, namely the multipartite entanglement is set by both UV and IR scales.  Hence, the entanglement in the strange metal contains UV-IR mixing.

A distinctive feature of the QFI computed here is that it is inversely proportional to temperature.  This is a consequence of the build-up of the density fluctuations seen experimentally in MEELS\cite{guo2024conformallyinvariantchargefluctuations} as the temperature is lowered.  Note, this behaviour is opposite of what we have found for a Fermi liquid.  Density fluctuations are expected to diverge at $T=0$ in a quantum critical system.  As the experiments are described by a conformally invariant theory, such a growth of the density fluctuations is expected. 

\begin{figure}
    \centering
    \includegraphics[width=\linewidth]{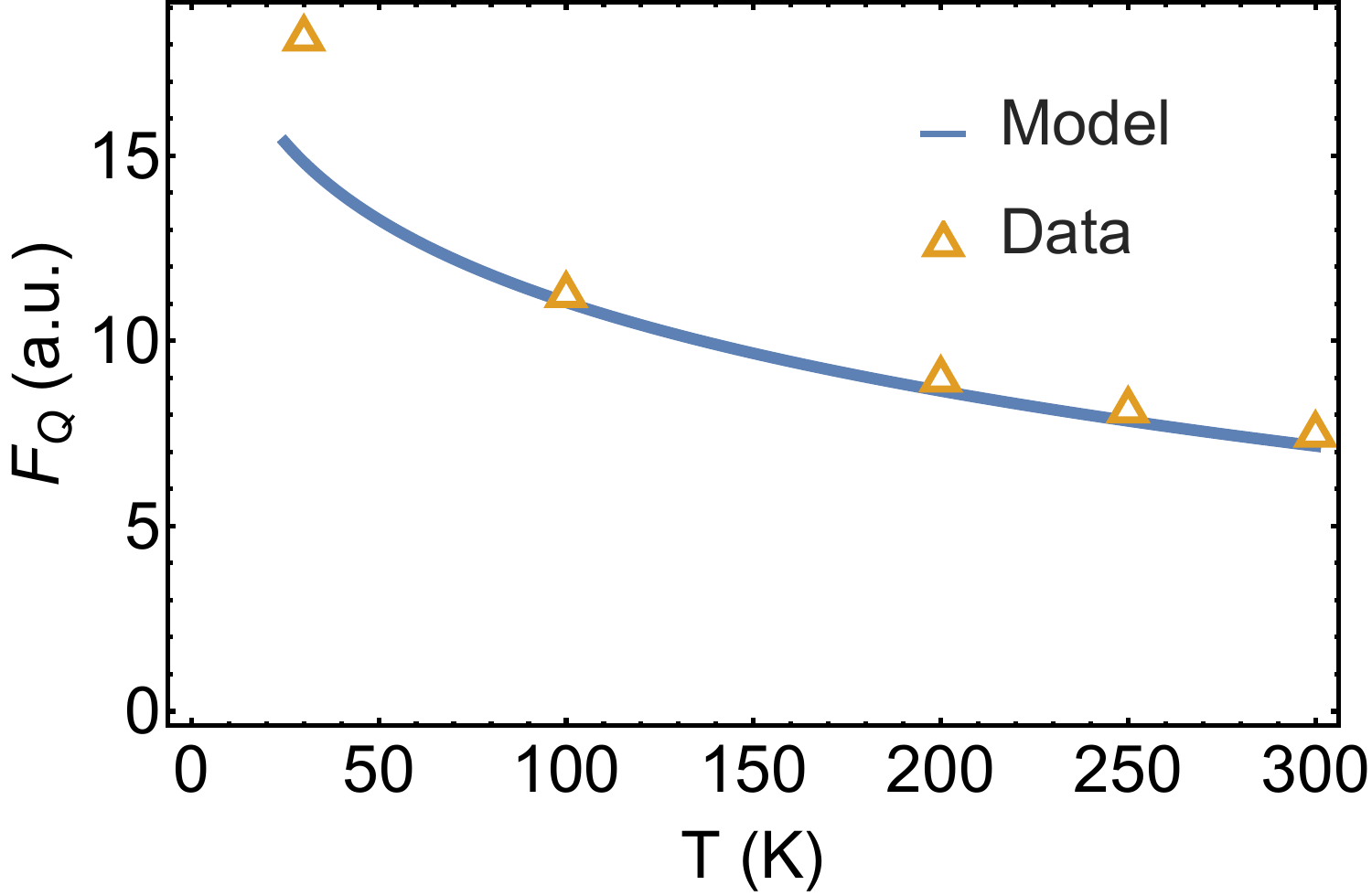}
    \caption{Comparison of the low-frequency contribution to the QFI determined from the experimental data (triangles) with the prediction from the conformal dynamical susceptibility (solid line) of Eq.~\eqref{cftf}. Both the experimental data and conformal dynamical susceptibility were integrated to $\omega=150eV\lesssim\omega_g$}
    \label{fig3}
\end{figure}

Finally, Fig.~\ref{fig3} shows a comparison of the low-frequency contribution to the QFI obtained from Eq. (\ref{cftf}) and a direct computation of Eq. (\ref{eq:qfidef}) from the experimental data, integrated to $\omega=150meV\lesssim\omega_g$. Note that experimentally, superconductivity onsets below $T\approx 90K$, below which the conformally-invariant model becomes inapplicable.  The steady increase in the QFI at low temperature is in sharp contrast with the Fermi liquid result. Although the high-resolution measurements allowing for computation of the QFI were performed only up to $150meV$, the approximate temperature independence of the susceptibility above this scale implies that the QFI for the strange metal is indeed dominated by entanglement due to UV-IR mixing as in Eq.~\eqref{uvir}.

\section{Conclusion}

We have shown here that the QFI is determined by the phase space for scattering. 
The $a+bT^2$ temperature dependence of the scattering phase space of a Fermi liquid explains the QFI we obtained in Fig. (\ref{fig1}) from Linhard RPA.  Contrastly, in the strange metal, the QFI increases as the temperature decreases.  The growth of such density fluctuations is consistent with approach to a $T=0$ critical point.  The constant at $T=0$, namely, $\omega_g^{2\Delta}$ suggests  that the strange metal has a multi-partite 
 entanglement spectrum that encodes physics of Mott insulation that arises entirely from spectral weight transfer. Fermi liquids lack such physics as Fig.~\ref{fig1} attests.  As the asymptotic value of the low temperature QFI in the strange metal is governed by the constant $\omega_g^{2\Delta}$, entanglement of the low-energy degrees of freedom is tethered to a UV scale. Such a temperature independence at low-T is consistent with the $T^{-\Delta_Q/z}$ scaling of the QFI only if $z=\infty$, a further indication that the underlying physics of the strange metal is governed by local quantum criticality.  Our analysis here is based entirely on the density-density susceptibility. While for non-interacting electrons, the density and spin susceptibilities are congruent QFIs\cite{si3}, the result for strongly correlated quantum matter remains an open question.
 
 
 Of course, UV-IR mixing in the strange metal is not surprising given that the cuprates are doped Mott insulators which exhibit both static and dynamical spectral weight transfer\cite{sawatzky,phillipsrmp}.  Such dynamical mixing between the lower and upper Hubbard bands signifies that the entanglement that ensues is not just a low-energy property, as the $\omega_g^{2\Delta}$ dependence of Eq.~\eqref{uvir} attests.  In essence, the entanglement in the QFI is a direct manifestation of Mottness\cite{phillipsrmp}, the intrinsic physics in Mott systems that has nothing to do with ordering. We thus speculate that the pseudogap sets the natural scale for the cutoff $\omega_g$.  Since this work illustrates that there is a marked departure from the Fermi liquid behavior in terms of the QFI, focusing on the entanglement as the root cause of T-linear resistivity\cite{hussey} seems like a reasonable starting point.   Our results appear to be consistent with the interpretation of the charge response as due to an emergent charge 2e boson arising from integrating out the upper Hubbard band, as argued in Ref.~\cite{2eprl}. Since that charge 2e boson has no inherent dynamics, the ultra-local nature of the charge response has a natural origin from the mixing between the two Hubbard bands. In fact, the dielectric response~\cite{phillipsexp} computed from the charge 2e boson theory shows a  peak and flat region at low energies that resembles the experimental data.  The peak corresponds to a bound sate of the charge 2e boson and a hole thereby leading to a depletion (pseudogap) of the spectral weight signified by zeros of the single-particle Green function. Our results raise the question of whether this charge 2e bosonic reduction of the Hubbard model\cite{2eprl} can be formulated as an ultra-local conformal theory in which the boson has the extremely low dimension $\Delta=0.05$. Perhaps only in $d=0+1$ dimensions can $\Delta=0.05$ be consistent with both unitarity and energy conservation~\cite{poland}. We defer a full exploration of this theoretical phenomenology to future work.

{\bf Acknowledgements.} 
We thank Jinchao Zhao for characteristically level-headed remarks.  This work was primarily supported by the Center for Quantum Sensing and Quantum Materials, an Energy Frontier Research Center funded by the U.S. Department of Energy (DOE), Office of Science, Basic Energy Sciences (BES), under award DE-SC0021238. Growth of Bi-2212 single crystals was supported by DOE Grant No. DE-SC0012704. P.A. gratefully acknowledges additional support from the EPiQS program of the Gordon and Betty Moore Foundation, grant GBMF9452.  
D.B. was supported in part by the A.C. Anderson Undergraduate Research Scholar Award, Department of Physics, University of Illinois Urbana-Champaign.
B. B. received additional support for theoretical development from the
National Science Foundation under  No. DMR-1945058. PWP also acknowledges NSF DMR-2111379 for partial funding.

\bibliographystyle{apsrev4-2}

\begin{thebibliography}{29}%
\makeatletter
\providecommand \@ifxundefined [1]{%
 \@ifx{#1\undefined}
}%
\providecommand \@ifnum [1]{%
 \ifnum #1\expandafter \@firstoftwo
 \else \expandafter \@secondoftwo
 \fi
}%
\providecommand \@ifx [1]{%
 \ifx #1\expandafter \@firstoftwo
 \else \expandafter \@secondoftwo
 \fi
}%
\providecommand \natexlab [1]{#1}%
\providecommand \enquote  [1]{``#1''}%
\providecommand \bibnamefont  [1]{#1}%
\providecommand \bibfnamefont [1]{#1}%
\providecommand \citenamefont [1]{#1}%
\providecommand \href@noop [0]{\@secondoftwo}%
\providecommand \href [0]{\begingroup \@sanitize@url \@href}%
\providecommand \@href[1]{\@@startlink{#1}\@@href}%
\providecommand \@@href[1]{\endgroup#1\@@endlink}%
\providecommand \@sanitize@url [0]{\catcode `\\12\catcode `\$12\catcode `\&12\catcode `\#12\catcode `\^12\catcode `\_12\catcode `\%12\relax}%
\providecommand \@@startlink[1]{}%
\providecommand \@@endlink[0]{}%
\providecommand \url  [0]{\begingroup\@sanitize@url \@url }%
\providecommand \@url [1]{\endgroup\@href {#1}{\urlprefix }}%
\providecommand \urlprefix  [0]{URL }%
\providecommand \Eprint [0]{\href }%
\providecommand \doibase [0]{https://doi.org/}%
\providecommand \selectlanguage [0]{\@gobble}%
\providecommand \bibinfo  [0]{\@secondoftwo}%
\providecommand \bibfield  [0]{\@secondoftwo}%
\providecommand \translation [1]{[#1]}%
\providecommand \BibitemOpen [0]{}%
\providecommand \bibitemStop [0]{}%
\providecommand \bibitemNoStop [0]{.\EOS\space}%
\providecommand \EOS [0]{\spacefactor3000\relax}%
\providecommand \BibitemShut  [1]{\csname bibitem#1\endcsname}%
\let\auto@bib@innerbib\@empty
\bibitem [{\citenamefont {Zaanen}(2019)}]{zaanenscipost}%
  \BibitemOpen
  \bibfield  {author} {\bibinfo {author} {\bibfnamefont {J.}~\bibnamefont {Zaanen}},\ }\href {https://doi.org/10.21468/SciPostPhys.6.5.061} {\bibfield  {journal} {\bibinfo  {journal} {SciPost Phys.}\ }\textbf {\bibinfo {volume} {6}},\ \bibinfo {pages} {061} (\bibinfo {year} {2019})}\BibitemShut {NoStop}%
\bibitem [{\citenamefont {Faulkner}\ \emph {et~al.}(2011{\natexlab{a}})\citenamefont {Faulkner}, \citenamefont {Liu}, \citenamefont {McGreevy},\ and\ \citenamefont {Vegh}}]{faulkner1}%
  \BibitemOpen
  \bibfield  {author} {\bibinfo {author} {\bibfnamefont {T.}~\bibnamefont {Faulkner}}, \bibinfo {author} {\bibfnamefont {H.}~\bibnamefont {Liu}}, \bibinfo {author} {\bibfnamefont {J.}~\bibnamefont {McGreevy}},\ and\ \bibinfo {author} {\bibfnamefont {D.}~\bibnamefont {Vegh}},\ }\href@noop {} {\bibfield  {journal} {\bibinfo  {journal} {Phys. Rev. D}\ }\textbf {\bibinfo {volume} {83}},\ \bibinfo {pages} {125002} (\bibinfo {year} {2011}{\natexlab{a}})}\BibitemShut {NoStop}%
\bibitem [{\citenamefont {Faulkner}\ \emph {et~al.}(2011{\natexlab{b}})\citenamefont {Faulkner}, \citenamefont {Horowitz},\ and\ \citenamefont {Roberts}}]{faulkner2}%
  \BibitemOpen
  \bibfield  {author} {\bibinfo {author} {\bibfnamefont {T.}~\bibnamefont {Faulkner}}, \bibinfo {author} {\bibfnamefont {G.~T.}\ \bibnamefont {Horowitz}},\ and\ \bibinfo {author} {\bibfnamefont {M.~M.}\ \bibnamefont {Roberts}},\ }\href@noop {} {\bibfield  {journal} {\bibinfo  {journal} {JHEP}\ }\textbf {\bibinfo {volume} {04}},\ \bibinfo {pages} {051}}\BibitemShut {NoStop}%
\bibitem [{\citenamefont {Guo}\ \emph {et~al.}(2024)\citenamefont {Guo}, \citenamefont {Chen}, \citenamefont {Hoveyda-Marashi}, \citenamefont {Bettler}, \citenamefont {Chaudhuri}, \citenamefont {Kengle}, \citenamefont {Schneeloch}, \citenamefont {Zhang}, \citenamefont {Gu}, \citenamefont {Chiang}, \citenamefont {Tsvelik}, \citenamefont {Faulkner}, \citenamefont {Phillips},\ and\ \citenamefont {Abbamonte}}]{guo2024conformallyinvariantchargefluctuations}%
  \BibitemOpen
  \bibfield  {author} {\bibinfo {author} {\bibfnamefont {X.}~\bibnamefont {Guo}}, \bibinfo {author} {\bibfnamefont {J.}~\bibnamefont {Chen}}, \bibinfo {author} {\bibfnamefont {F.}~\bibnamefont {Hoveyda-Marashi}}, \bibinfo {author} {\bibfnamefont {S.~L.}\ \bibnamefont {Bettler}}, \bibinfo {author} {\bibfnamefont {D.}~\bibnamefont {Chaudhuri}}, \bibinfo {author} {\bibfnamefont {C.~S.}\ \bibnamefont {Kengle}}, \bibinfo {author} {\bibfnamefont {J.~A.}\ \bibnamefont {Schneeloch}}, \bibinfo {author} {\bibfnamefont {R.}~\bibnamefont {Zhang}}, \bibinfo {author} {\bibfnamefont {G.}~\bibnamefont {Gu}}, \bibinfo {author} {\bibfnamefont {T.-C.}\ \bibnamefont {Chiang}}, \bibinfo {author} {\bibfnamefont {A.~M.}\ \bibnamefont {Tsvelik}}, \bibinfo {author} {\bibfnamefont {T.}~\bibnamefont {Faulkner}}, \bibinfo {author} {\bibfnamefont {P.~W.}\ \bibnamefont {Phillips}},\ and\ \bibinfo {author} {\bibfnamefont {P.}~\bibnamefont {Abbamonte}},\ }\href {https://arxiv.org/abs/2411.11164} {\bibinfo {title} {Conformally invariant
  charge fluctuations in a strange metal}} (\bibinfo {year} {2024}),\ \Eprint {https://arxiv.org/abs/2411.11164} {arXiv:2411.11164 [cond-mat.str-el]} \BibitemShut {NoStop}%
\bibitem [{\citenamefont {Mitrano}\ \emph {et~al.}()\citenamefont {Mitrano}, \citenamefont {Husain}, \citenamefont {Vig}, \citenamefont {Kogar}, \citenamefont {Rak}, \citenamefont {Rubeck}, \citenamefont {Schmalian}, \citenamefont {Uchoa}, \citenamefont {Schneeloch}, \citenamefont {Zhong}, \citenamefont {Gu},\ and\ \citenamefont {Abbamonte}}]{mitranoAnomalousDensityFluctuations2018}%
  \BibitemOpen
  \bibfield  {author} {\bibinfo {author} {\bibfnamefont {M.}~\bibnamefont {Mitrano}}, \bibinfo {author} {\bibfnamefont {A.~A.}\ \bibnamefont {Husain}}, \bibinfo {author} {\bibfnamefont {S.}~\bibnamefont {Vig}}, \bibinfo {author} {\bibfnamefont {A.}~\bibnamefont {Kogar}}, \bibinfo {author} {\bibfnamefont {M.~S.}\ \bibnamefont {Rak}}, \bibinfo {author} {\bibfnamefont {S.~I.}\ \bibnamefont {Rubeck}}, \bibinfo {author} {\bibfnamefont {J.}~\bibnamefont {Schmalian}}, \bibinfo {author} {\bibfnamefont {B.}~\bibnamefont {Uchoa}}, \bibinfo {author} {\bibfnamefont {J.}~\bibnamefont {Schneeloch}}, \bibinfo {author} {\bibfnamefont {R.}~\bibnamefont {Zhong}}, \bibinfo {author} {\bibfnamefont {G.~D.}\ \bibnamefont {Gu}},\ and\ \bibinfo {author} {\bibfnamefont {P.}~\bibnamefont {Abbamonte}},\ }\href {https://doi.org/10.1073/pnas.1721495115} {\bibfield  {journal} {\bibinfo  {journal} {Proceedings of the National Academy of Sciences}\ }\textbf {\bibinfo {volume} {115}},\ \bibinfo {pages} {5392}}\BibitemShut {NoStop}%
\bibitem [{\citenamefont {Aronson}\ \emph {et~al.}(1997)\citenamefont {Aronson}, \citenamefont {Maple}, \citenamefont {de~Sa}, \citenamefont {Tsvelik},\ and\ \citenamefont {Osborn}}]{Tsvelik1997}%
  \BibitemOpen
  \bibfield  {author} {\bibinfo {author} {\bibfnamefont {M.~C.}\ \bibnamefont {Aronson}}, \bibinfo {author} {\bibfnamefont {M.~B.}\ \bibnamefont {Maple}}, \bibinfo {author} {\bibfnamefont {P.}~\bibnamefont {de~Sa}}, \bibinfo {author} {\bibfnamefont {A.~M.}\ \bibnamefont {Tsvelik}},\ and\ \bibinfo {author} {\bibfnamefont {R.}~\bibnamefont {Osborn}},\ }\href@noop {} {\bibfield  {journal} {\bibinfo  {journal} {Europhys. Lett.}\ }\textbf {\bibinfo {volume} {40}},\ \bibinfo {pages} {245} (\bibinfo {year} {1997})}\BibitemShut {NoStop}%
\bibitem [{\citenamefont {Varma}\ \emph {et~al.}(1989)\citenamefont {Varma}, \citenamefont {Littlewood}, \citenamefont {Schmitt-Rink}, \citenamefont {Abrahams},\ and\ \citenamefont {Ruckenstein}}]{mfl}%
  \BibitemOpen
  \bibfield  {author} {\bibinfo {author} {\bibfnamefont {C.~M.}\ \bibnamefont {Varma}}, \bibinfo {author} {\bibfnamefont {P.~B.}\ \bibnamefont {Littlewood}}, \bibinfo {author} {\bibfnamefont {S.}~\bibnamefont {Schmitt-Rink}}, \bibinfo {author} {\bibfnamefont {E.}~\bibnamefont {Abrahams}},\ and\ \bibinfo {author} {\bibfnamefont {A.~E.}\ \bibnamefont {Ruckenstein}},\ }\href@noop {} {\bibfield  {journal} {\bibinfo  {journal} {Phys. Rev. Lett.}\ }\textbf {\bibinfo {volume} {63}},\ \bibinfo {pages} {1996} (\bibinfo {year} {1989})}\BibitemShut {NoStop}%
\bibitem [{\citenamefont {Legros}\ \emph {et~al.}(2019)\citenamefont {Legros}, \citenamefont {Benhabib}, \citenamefont {Tabis}, \citenamefont {Lalibert{\'e}}, \citenamefont {Dion}, \citenamefont {Lizaire}, \citenamefont {Vignolle}, \citenamefont {Vignolles}, \citenamefont {Raffy}, \citenamefont {Li}, \citenamefont {Auban-Senzier}, \citenamefont {Doiron-Leyraud}, \citenamefont {Fournier}, \citenamefont {Colson}, \citenamefont {Taillefer},\ and\ \citenamefont {Proust}}]{legros}%
  \BibitemOpen
  \bibfield  {author} {\bibinfo {author} {\bibfnamefont {A.}~\bibnamefont {Legros}}, \bibinfo {author} {\bibfnamefont {S.}~\bibnamefont {Benhabib}}, \bibinfo {author} {\bibfnamefont {W.}~\bibnamefont {Tabis}}, \bibinfo {author} {\bibfnamefont {F.}~\bibnamefont {Lalibert{\'e}}}, \bibinfo {author} {\bibfnamefont {M.}~\bibnamefont {Dion}}, \bibinfo {author} {\bibfnamefont {M.}~\bibnamefont {Lizaire}}, \bibinfo {author} {\bibfnamefont {B.}~\bibnamefont {Vignolle}}, \bibinfo {author} {\bibfnamefont {D.}~\bibnamefont {Vignolles}}, \bibinfo {author} {\bibfnamefont {H.}~\bibnamefont {Raffy}}, \bibinfo {author} {\bibfnamefont {Z.~Z.}\ \bibnamefont {Li}}, \bibinfo {author} {\bibfnamefont {P.}~\bibnamefont {Auban-Senzier}}, \bibinfo {author} {\bibfnamefont {N.}~\bibnamefont {Doiron-Leyraud}}, \bibinfo {author} {\bibfnamefont {P.}~\bibnamefont {Fournier}}, \bibinfo {author} {\bibfnamefont {D.}~\bibnamefont {Colson}}, \bibinfo {author} {\bibfnamefont {L.}~\bibnamefont {Taillefer}},\ and\ \bibinfo {author} {\bibfnamefont
  {C.}~\bibnamefont {Proust}},\ }\href {https://doi.org/10.1038/s41567-018-0334-2} {\bibfield  {journal} {\bibinfo  {journal} {Nature Physics}\ }\textbf {\bibinfo {volume} {15}},\ \bibinfo {pages} {142} (\bibinfo {year} {2019})}\BibitemShut {NoStop}%
\bibitem [{\citenamefont {Son}\ and\ \citenamefont {Starinets}(2007)}]{etas}%
  \BibitemOpen
  \bibfield  {author} {\bibinfo {author} {\bibfnamefont {D.~T.}\ \bibnamefont {Son}}\ and\ \bibinfo {author} {\bibfnamefont {A.~O.}\ \bibnamefont {Starinets}},\ }\href {https://doi.org/https://doi.org/10.1146/annurev.nucl.57.090506.123120} {\bibfield  {journal} {\bibinfo  {journal} {Annual Review of Nuclear and Particle Science}\ }\textbf {\bibinfo {volume} {57}},\ \bibinfo {pages} {95} (\bibinfo {year} {2007})}\BibitemShut {NoStop}%
\bibitem [{\citenamefont {Srednicki}(1993)}]{srednicki}%
  \BibitemOpen
  \bibfield  {author} {\bibinfo {author} {\bibfnamefont {M.}~\bibnamefont {Srednicki}},\ }\href {https://doi.org/10.1103/PhysRevLett.71.666} {\bibfield  {journal} {\bibinfo  {journal} {Phys. Rev. Lett.}\ }\textbf {\bibinfo {volume} {71}},\ \bibinfo {pages} {666} (\bibinfo {year} {1993})}\BibitemShut {NoStop}%
\bibitem [{\citenamefont {Guo}\ \emph {et~al.}(2022)\citenamefont {Guo}, \citenamefont {Jia}, \citenamefont {Li},\ and\ \citenamefont {Huang}}]{Guo_2022}%
  \BibitemOpen
  \bibfield  {author} {\bibinfo {author} {\bibfnamefont {Y.}~\bibnamefont {Guo}}, \bibinfo {author} {\bibfnamefont {Y.}~\bibnamefont {Jia}}, \bibinfo {author} {\bibfnamefont {X.}~\bibnamefont {Li}},\ and\ \bibinfo {author} {\bibfnamefont {L.}~\bibnamefont {Huang}},\ }\href {https://doi.org/10.1088/1751-8121/ac5649} {\bibfield  {journal} {\bibinfo  {journal} {Journal of Physics A: Mathematical and Theoretical}\ }\textbf {\bibinfo {volume} {55}},\ \bibinfo {pages} {145303} (\bibinfo {year} {2022})}\BibitemShut {NoStop}%
\bibitem [{\citenamefont {Horodecki}\ \emph {et~al.}(2009)\citenamefont {Horodecki}, \citenamefont {Horodecki}, \citenamefont {Horodecki},\ and\ \citenamefont {Horodecki}}]{multi1}%
  \BibitemOpen
  \bibfield  {author} {\bibinfo {author} {\bibfnamefont {R.}~\bibnamefont {Horodecki}}, \bibinfo {author} {\bibfnamefont {P.}~\bibnamefont {Horodecki}}, \bibinfo {author} {\bibfnamefont {M.}~\bibnamefont {Horodecki}},\ and\ \bibinfo {author} {\bibfnamefont {K.}~\bibnamefont {Horodecki}},\ }\href {https://doi.org/10.1103/RevModPhys.81.865} {\bibfield  {journal} {\bibinfo  {journal} {Rev. Mod. Phys.}\ }\textbf {\bibinfo {volume} {81}},\ \bibinfo {pages} {865} (\bibinfo {year} {2009})}\BibitemShut {NoStop}%
\bibitem [{\citenamefont {T\'oth}(2012)}]{Toth}%
  \BibitemOpen
  \bibfield  {author} {\bibinfo {author} {\bibfnamefont {G.}~\bibnamefont {T\'oth}},\ }\href {https://doi.org/10.1103/PhysRevA.85.022322} {\bibfield  {journal} {\bibinfo  {journal} {Phys. Rev. A}\ }\textbf {\bibinfo {volume} {85}},\ \bibinfo {pages} {022322} (\bibinfo {year} {2012})}\BibitemShut {NoStop}%
\bibitem [{\citenamefont {Hyllus}\ \emph {et~al.}(2012)\citenamefont {Hyllus}, \citenamefont {Laskowski}, \citenamefont {Krischek}, \citenamefont {Schwemmer}, \citenamefont {Wieczorek}, \citenamefont {Weinfurter}, \citenamefont {Pezz\'e},\ and\ \citenamefont {Smerzi}}]{hyllus}%
  \BibitemOpen
  \bibfield  {author} {\bibinfo {author} {\bibfnamefont {P.}~\bibnamefont {Hyllus}}, \bibinfo {author} {\bibfnamefont {W.}~\bibnamefont {Laskowski}}, \bibinfo {author} {\bibfnamefont {R.}~\bibnamefont {Krischek}}, \bibinfo {author} {\bibfnamefont {C.}~\bibnamefont {Schwemmer}}, \bibinfo {author} {\bibfnamefont {W.}~\bibnamefont {Wieczorek}}, \bibinfo {author} {\bibfnamefont {H.}~\bibnamefont {Weinfurter}}, \bibinfo {author} {\bibfnamefont {L.}~\bibnamefont {Pezz\'e}},\ and\ \bibinfo {author} {\bibfnamefont {A.}~\bibnamefont {Smerzi}},\ }\href {https://doi.org/10.1103/PhysRevA.85.022321} {\bibfield  {journal} {\bibinfo  {journal} {Phys. Rev. A}\ }\textbf {\bibinfo {volume} {85}},\ \bibinfo {pages} {022321} (\bibinfo {year} {2012})}\BibitemShut {NoStop}%
\bibitem [{\citenamefont {Scheie}\ \emph {et~al.}(2021)\citenamefont {Scheie}, \citenamefont {Laurell}, \citenamefont {Samarakoon}, \citenamefont {Lake}, \citenamefont {Nagler}, \citenamefont {Granroth}, \citenamefont {Okamoto}, \citenamefont {Alvarez},\ and\ \citenamefont {Tennant}}]{schiewitness}%
  \BibitemOpen
  \bibfield  {author} {\bibinfo {author} {\bibfnamefont {A.}~\bibnamefont {Scheie}}, \bibinfo {author} {\bibfnamefont {P.}~\bibnamefont {Laurell}}, \bibinfo {author} {\bibfnamefont {A.~M.}\ \bibnamefont {Samarakoon}}, \bibinfo {author} {\bibfnamefont {B.}~\bibnamefont {Lake}}, \bibinfo {author} {\bibfnamefont {S.~E.}\ \bibnamefont {Nagler}}, \bibinfo {author} {\bibfnamefont {G.~E.}\ \bibnamefont {Granroth}}, \bibinfo {author} {\bibfnamefont {S.}~\bibnamefont {Okamoto}}, \bibinfo {author} {\bibfnamefont {G.}~\bibnamefont {Alvarez}},\ and\ \bibinfo {author} {\bibfnamefont {D.~A.}\ \bibnamefont {Tennant}},\ }\href {https://doi.org/10.1103/PhysRevB.103.224434} {\bibfield  {journal} {\bibinfo  {journal} {Phys. Rev. B}\ }\textbf {\bibinfo {volume} {103}},\ \bibinfo {pages} {224434} (\bibinfo {year} {2021})}\BibitemShut {NoStop}%
\bibitem [{\citenamefont {Menon}\ \emph {et~al.}(2023)\citenamefont {Menon}, \citenamefont {Sherman}, \citenamefont {Dupont}, \citenamefont {Scheie}, \citenamefont {Tennant},\ and\ \citenamefont {Moore}}]{witness2}%
  \BibitemOpen
  \bibfield  {author} {\bibinfo {author} {\bibfnamefont {V.}~\bibnamefont {Menon}}, \bibinfo {author} {\bibfnamefont {N.~E.}\ \bibnamefont {Sherman}}, \bibinfo {author} {\bibfnamefont {M.}~\bibnamefont {Dupont}}, \bibinfo {author} {\bibfnamefont {A.~O.}\ \bibnamefont {Scheie}}, \bibinfo {author} {\bibfnamefont {D.~A.}\ \bibnamefont {Tennant}},\ and\ \bibinfo {author} {\bibfnamefont {J.~E.}\ \bibnamefont {Moore}},\ }\href {https://doi.org/10.1103/PhysRevB.107.054422} {\bibfield  {journal} {\bibinfo  {journal} {Phys. Rev. B}\ }\textbf {\bibinfo {volume} {107}},\ \bibinfo {pages} {054422} (\bibinfo {year} {2023})}\BibitemShut {NoStop}%

\bibitem [{\citenamefont {Fisher}(1925)}]{fisher}%
  \BibitemOpen
  \bibfield  {author} {\bibinfo {author} {\bibfnamefont {R.~A.}~\bibnamefont {Fisher}},\ }
  \href {https://doi.org/10.1017/S0305004100009580} {\bibfield  {journal} {\bibinfo  {journal} {Math. Proc. Camb. Philos. Soc.}\ }\textbf {\bibinfo {volume} {22}},\ \bibinfo {pages} {700–725} (\bibinfo {year} {1925})}\BibitemShut {NoStop}%

\bibitem [{\citenamefont {Braunstein}\ and\ \citenamefont {Caves}(1994)}]{bures}%
  \BibitemOpen
  \bibfield  {author} {\bibinfo {author} {\bibfnamefont {S.~L.}\ \bibnamefont {Braunstein}}\ and\ \bibinfo {author} {\bibfnamefont {C.~M.}\ \bibnamefont {Caves}},\ }\href {https://doi.org/10.1103/PhysRevLett.72.3439} {\bibfield  {journal} {\bibinfo  {journal} {Phys. Rev. Lett.}\ }\textbf {\bibinfo {volume} {72}},\ \bibinfo {pages} {3439} (\bibinfo {year} {1994})}\BibitemShut {NoStop}%
\bibitem [{\citenamefont {Provost}\ and\ \citenamefont {Vallee}(1980)}]{provost1980riemannian}%
  \BibitemOpen
  \bibfield  {author} {\bibinfo {author} {\bibfnamefont {J.}~\bibnamefont {Provost}}\ and\ \bibinfo {author} {\bibfnamefont {G.}~\bibnamefont {Vallee}},\ }\href@noop {} {\bibfield  {journal} {\bibinfo  {journal} {Communications in Mathematical Physics}\ }\textbf {\bibinfo {volume} {76}},\ \bibinfo {pages} {289} (\bibinfo {year} {1980})}\BibitemShut {NoStop}%
\bibitem [{\citenamefont {Hauke}\ \emph {et~al.}()\citenamefont {Hauke}, \citenamefont {Heyl}, \citenamefont {Tagliacozzo},\ and\ \citenamefont {Zoller}}]{haukeMeasuringMultipartiteEntanglement2016}%
  \BibitemOpen
  \bibfield  {author} {\bibinfo {author} {\bibfnamefont {P.}~\bibnamefont {Hauke}}, \bibinfo {author} {\bibfnamefont {M.}~\bibnamefont {Heyl}}, \bibinfo {author} {\bibfnamefont {L.}~\bibnamefont {Tagliacozzo}},\ and\ \bibinfo {author} {\bibfnamefont {P.}~\bibnamefont {Zoller}},\ }\href {https://doi.org/10.1038/nphys3700} {\bibfield  {journal} {\bibinfo  {journal} {Nature Physics}\ }\textbf {\bibinfo {volume} {12}},\ \bibinfo {pages} {778}}\BibitemShut {NoStop}%
\bibitem [{\citenamefont {Eskes}\ \emph {et~al.}(1991)\citenamefont {Eskes}, \citenamefont {Meinders},\ and\ \citenamefont {Sawatzky}}]{sawatzky}%
  \BibitemOpen
  \bibfield  {author} {\bibinfo {author} {\bibfnamefont {H.}~\bibnamefont {Eskes}}, \bibinfo {author} {\bibfnamefont {M.~B.~J.}\ \bibnamefont {Meinders}},\ and\ \bibinfo {author} {\bibfnamefont {G.~A.}\ \bibnamefont {Sawatzky}},\ }\href {https://doi.org/10.1103/PhysRevLett.67.1035} {\bibfield  {journal} {\bibinfo  {journal} {Phys. Rev. Lett.}\ }\textbf {\bibinfo {volume} {67}},\ \bibinfo {pages} {1035} (\bibinfo {year} {1991})}\BibitemShut {NoStop}%
\bibitem [{\citenamefont {Phillips}(2010)}]{phillipsrmp}%
  \BibitemOpen
  \bibfield  {author} {\bibinfo {author} {\bibfnamefont {P.}~\bibnamefont {Phillips}},\ }\href {https://doi.org/10.1103/RevModPhys.82.1719} {\bibfield  {journal} {\bibinfo  {journal} {Rev. Mod. Phys.}\ }\textbf {\bibinfo {volume} {82}},\ \bibinfo {pages} {1719} (\bibinfo {year} {2010})}\BibitemShut {NoStop}%
\bibitem [{\citenamefont {Bałut}\ \emph {et~al.}()\citenamefont {Bałut}, \citenamefont {Bradlyn},\ and\ \citenamefont {Abbamonte}}]{balutQuantumEntanglementQuantum2024}%
  \BibitemOpen
  \bibfield  {author} {\bibinfo {author} {\bibfnamefont {D.}~\bibnamefont {Bałut}}, \bibinfo {author} {\bibfnamefont {B.}~\bibnamefont {Bradlyn}},\ and\ \bibinfo {author} {\bibfnamefont {P.}~\bibnamefont {Abbamonte}},\ }\href {https://doi.org/10.48550/arXiv.2409.15583} {\bibinfo {title} {Quantum entanglement and quantum geometry measured with inelastic {{X-ray}} scattering}},\ \Eprint {https://arxiv.org/abs/2409.15583} {2409.15583} \BibitemShut {NoStop}%
\bibitem [{\citenamefont {Mazza}\ \emph {et~al.}(2024)\citenamefont {Mazza}, \citenamefont {Biswas}, \citenamefont {Yan}, \citenamefont {Prokofiev}, \citenamefont {Steffens}, \citenamefont {Si}, \citenamefont {Assaad},\ and\ \citenamefont {Paschen}}]{mazza}%
  \BibitemOpen
  \bibfield  {author} {\bibinfo {author} {\bibfnamefont {F.}~\bibnamefont {Mazza}}, \bibinfo {author} {\bibfnamefont {S.}~\bibnamefont {Biswas}}, \bibinfo {author} {\bibfnamefont {X.}~\bibnamefont {Yan}}, \bibinfo {author} {\bibfnamefont {A.}~\bibnamefont {Prokofiev}}, \bibinfo {author} {\bibfnamefont {P.}~\bibnamefont {Steffens}}, \bibinfo {author} {\bibfnamefont {Q.}~\bibnamefont {Si}}, \bibinfo {author} {\bibfnamefont {F.~F.}\ \bibnamefont {Assaad}},\ and\ \bibinfo {author} {\bibfnamefont {S.}~\bibnamefont {Paschen}},\ }\href {https://arxiv.org/abs/2403.12779} {\bibinfo {title} {Quantum fisher information in a strange metal}} (\bibinfo {year} {2024}),\ \Eprint {https://arxiv.org/abs/2403.12779} {arXiv:2403.12779 [cond-mat.str-el]} \BibitemShut {NoStop}%
\bibitem [{\citenamefont {Fetter}\ and\ \citenamefont {Walecka}(1971)}]{FW}%
  \BibitemOpen
  \bibfield  {author} {\bibinfo {author} {\bibfnamefont {A.~L.}\ \bibnamefont {Fetter}}\ and\ \bibinfo {author} {\bibfnamefont {J.~D.}\ \bibnamefont {Walecka}},\ }\href@noop {} {\emph {\bibinfo {title} {Quantum Theory of Many-Particle Systems}}}\ (\bibinfo  {publisher} {McGraw-Hill},\ \bibinfo {address} {Boston},\ \bibinfo {year} {1971})\BibitemShut {NoStop}%
\bibitem [{\citenamefont {Phillips}(2012)}]{phillipsbook}%
  \BibitemOpen
  \bibfield  {author} {\bibinfo {author} {\bibfnamefont {P.}~\bibnamefont {Phillips}},\ }\href@noop {} {\emph {\bibinfo {title} {Advanced Solid State Physics}}},\ \bibinfo {edition} {2nd}\ ed.\ (\bibinfo  {publisher} {Cambridge University Press},\ \bibinfo {year} {2012})\BibitemShut {NoStop}%
\bibitem [{\citenamefont {Phillips}\ \emph {et~al.}(2022)\citenamefont {Phillips}, \citenamefont {Hussey},\ and\ \citenamefont {Abbamonte}}]{hussey}%
  \BibitemOpen
  \bibfield  {author} {\bibinfo {author} {\bibfnamefont {P.~W.}\ \bibnamefont {Phillips}}, \bibinfo {author} {\bibfnamefont {N.~E.}\ \bibnamefont {Hussey}},\ and\ \bibinfo {author} {\bibfnamefont {P.}~\bibnamefont {Abbamonte}},\ }\href@noop {} {\bibfield  {journal} {\bibinfo  {journal} {Science}\ }\textbf {\bibinfo {volume} {377}},\ \bibinfo {pages} {eabh4273} (\bibinfo {year} {2022})}\BibitemShut {NoStop}%
\bibitem [{\citenamefont {Leigh}\ \emph {et~al.}(2007)\citenamefont {Leigh}, \citenamefont {Phillips},\ and\ \citenamefont {Choy}}]{2eprl}%
  \BibitemOpen
  \bibfield  {author} {\bibinfo {author} {\bibfnamefont {R.~G.}\ \bibnamefont {Leigh}}, \bibinfo {author} {\bibfnamefont {P.}~\bibnamefont {Phillips}},\ and\ \bibinfo {author} {\bibfnamefont {T.-P.}\ \bibnamefont {Choy}},\ }\href {https://doi.org/10.1103/PhysRevLett.99.046404} {\bibfield  {journal} {\bibinfo  {journal} {Phys. Rev. Lett.}\ }\textbf {\bibinfo {volume} {99}},\ \bibinfo {pages} {046404} (\bibinfo {year} {2007})}\BibitemShut {NoStop}%
\bibitem [{\citenamefont {Choy}\ \emph {et~al.}(2008)\citenamefont {Choy}, \citenamefont {Leigh},\ and\ \citenamefont {Phillips}}]{phillipsexp}%
  \BibitemOpen
  \bibfield  {author} {\bibinfo {author} {\bibfnamefont {T.-P.}\ \bibnamefont {Choy}}, \bibinfo {author} {\bibfnamefont {R.~G.}\ \bibnamefont {Leigh}},\ and\ \bibinfo {author} {\bibfnamefont {P.}~\bibnamefont {Phillips}},\ }\href {https://doi.org/10.1103/PhysRevB.77.104524} {\bibfield  {journal} {\bibinfo  {journal} {Phys. Rev. B}\ }\textbf {\bibinfo {volume} {77}},\ \bibinfo {pages} {104524} (\bibinfo {year} {2008})}\BibitemShut {NoStop}%
\bibitem [{\citenamefont {Poland}\ \emph {et~al.}(2019)\citenamefont {Poland}, \citenamefont {Rychkov},\ and\ \citenamefont {Vichi}}]{poland}%
  \BibitemOpen
  \bibfield  {author} {\bibinfo {author} {\bibfnamefont {D.}~\bibnamefont {Poland}}, \bibinfo {author} {\bibfnamefont {S.}~\bibnamefont {Rychkov}},\ and\ \bibinfo {author} {\bibfnamefont {A.}~\bibnamefont {Vichi}},\ }\href {https://doi.org/10.1103/RevModPhys.91.015002} {\bibfield  {journal} {\bibinfo  {journal} {Rev. Mod. Phys.}\ }\textbf {\bibinfo {volume} {91}},\ \bibinfo {pages} {015002} (\bibinfo {year} {2019})}\BibitemShut {NoStop}%

\bibitem [{\citenamefont {Fang}\ \emph {et~al.}(2024)\citenamefont {Fang}, \citenamefont {Mahankali}, \citenamefont {Wang}, \citenamefont {Chen}, \citenamefont {Hu}, \citenamefont {Paschen},\ and\ \citenamefont {Si}}]{si2}%
  \BibitemOpen
  \bibfield  {author} {\bibinfo {author} {\bibfnamefont {Y.}~\bibnamefont {Fang}},  
  \bibinfo {author} {\bibfnamefont {M.}~\bibnamefont {Mahankali}},  
  \bibinfo {author} {\bibfnamefont {Y.}~\bibnamefont {Wang}},  
  \bibinfo {author} {\bibfnamefont {L.}~\bibnamefont {Chen}},  
  \bibinfo {author} {\bibfnamefont {H.}~\bibnamefont {Hu}},  
  \bibinfo {author} {\bibfnamefont {S.}~\bibnamefont {Paschen}},\ and\  
  \bibinfo {author} {\bibfnamefont {Q.}~\bibnamefont {Si}},\ }
  \href {https://doi.org/10.48550/arXiv.2402.18552} {\bibfield  {journal} {\bibinfo  {journal} {arXiv e-prints}\ }\textbf {\bibinfo {volume} {arXiv:2402.18552}} (\bibinfo {year} {2024})}\BibitemShut {NoStop}%

\bibitem [{\citenamefont {Mihaila}(2011)}]{Mihaila_2011}%
  \BibitemOpen
  \bibfield  {author} {\bibinfo {author} {\bibfnamefont {Bogdan}~\bibnamefont {Mihaila}},\ }
  \href {http://arxiv.org/abs/1111.5337} {\bibfield  {journal} {\bibinfo  {journal} {arXiv}\ }
  \textbf {\bibinfo {number} {1111.5337}},\  (\bibinfo {year} {2011})}\BibitemShut {NoStop}%

  \bibitem [{\citenamefont {Wang}\ \emph {et~al.}(2025)\citenamefont {Wang}, \citenamefont {Fang}, \citenamefont {Xie},\ and\ \citenamefont {Si}}]{si3}%
  \BibitemOpen
  \bibfield  {author} {\bibinfo {author} {\bibfnamefont {Y.}~\bibnamefont {Wang}},  
  \bibinfo {author} {\bibfnamefont {Y.}~\bibnamefont {Fang}},  
  \bibinfo {author} {\bibfnamefont {F.}~\bibnamefont {Xie}},\ and\  
  \bibinfo {author} {\bibfnamefont {Q.}~\bibnamefont {Si}},\ }
  \href {https://doi.org/10.48550/arXiv.2502.13958} {\bibfield  {journal} {\bibinfo  {journal} {arXiv e-prints}\ }\textbf {\bibinfo {volume} {arXiv:2502.13958}} (\bibinfo {year} {2025})}\BibitemShut {NoStop}%

\end{thebibliography}

\end{document}